\documentclass[acmsmall,nonacm,screen]{acmart}
\usepackage{subcaption} 
\usepackage{listings}
\usepackage[framemethod=TikZ]{mdframed}
\usepackage{fontawesome5}
\usepackage[linesnumbered,ruled]{algorithm2e}
\usepackage{multirow}
\usepackage{colortbl}
\usepackage{enumitem}

\SetCommentSty{textnormal}
\SetNlSty{textbf}{\rmfamily}{}

\newmdenv[
  linewidth=1pt,
  linecolor=black,
  backgroundcolor=gray!10,
  skipabove=10pt,
  skipbelow=10pt,
  roundcorner=5pt,
  innertopmargin=10pt,
  innerbottommargin=10pt,
  innerleftmargin=10pt,
  innerrightmargin=10pt,
  frametitlebackgroundcolor=gray!30!black,
  frametitlefont=\color{white}\bfseries,
]{promptbox}

\newcommand{\prompt}[2]{%
  \begin{promptbox}[frametitle={#1}]
  #2
  \end{promptbox}
}

\newmdenv[
  backgroundcolor=gray!20, 
  linecolor=black,          
  linewidth=1pt,            
  roundcorner=5pt,          
  skipabove=10pt,           
  skipbelow=10pt,           
  innerleftmargin=5pt,
  innerrightmargin=5pt,
  innertopmargin=5pt,
  innerbottommargin=5pt,
]{answerboxenv}

\newcommand{\answerbox}[2][gray!20]{%
  \begin{answerboxenv}[backgroundcolor=#1]%
  #2%
  \end{answerboxenv}%
}

\newcommand{\dashedline}{
  \noindent
  \leavevmode
  \leaders\hbox to 0.7em{\hss\textendash\hss}\hfill\kern0pt
}

\definecolor{codegreen}{rgb}{0,0.6,0}
\definecolor{codepurple}{rgb}{0.58,0,0.82}
\definecolor{codegray}{gray}{.9}
\definecolor{backcolour}{rgb}{0.95,0.95,0.92}
\definecolor{addgreen}{rgb}{0.086, 0.588, 0.349}
\definecolor{rmred}{rgb}{0.861, 0.129, 0.129} 

\lstdefinestyle{customgo}{
        language=go,
        numbers=left,    
        numbersep=4pt,
        morekeywords={exp256Modulus,Exp,toI256,Int,big.Int},
        moredelim=[is][\color{codegreen}]{!!}{!!}, 
        moredelim=[is][\color{rmred}]{??}{??}, 
}

\lstdefinestyle{gocomment}{
        language=go,
        numbers=left,    
        numbersep=4pt,
        moredelim=[is][\color{codegreen}]{!!}{!!}, 
        moredelim=[is][\color{rmred}]{??}{??}, 
}

\setlist{leftmargin=*}

\AtBeginDocument{%
  }

\acmDOI{10.1145/3728946}
\acmYear{2025}
\acmJournal{PACMSE}
\acmVolume{2}
\acmNumber{ISSTA}
\acmArticle{ISSTA069}
\acmMonth{7}
\received{2024-10-31}
\received[accepted]{2025-03-31}

\begin{document}

\newcommand{\framework}{\textsc{OpDiffer}}

\title{{\framework}: LLM-Assisted Opcode-Level Differential Testing of Ethereum Virtual Machine}

\author{Jie Ma}
\orcid{0009-0007-6311-520X}
\affiliation{%
  \institution{Beihang University}
  \city{Beijing}
  \country{China}
}
\affiliation{%
  \institution{Zhongguancun Laboratory}
  \city{Beijing}
  \country{China}
}
\email{majie2023@buaa.edu.cn}

\author{Ningyu He}
\orcid{0000-0002-9980-7298}
\affiliation{%
  \institution{The Hong Kong Polytechnic University}
  \city{Hong Kong}
  \country{China}
}
\email{ningyu.he@pku.edu.cn}

\author{Jinwen Xi}
\orcid{0000-0002-7504-3457}
\affiliation{%
  \institution{Zhongguancun Laboratory}
  \city{Beijing}
  \country{China}
}
\email{xijw@zgclab.edu.cn}

\author{Mingzhe Xing}
\orcid{0000-0002-2065-9852}
\affiliation{%
  \institution{Zhongguancun Laboratory}
  \city{Beijing}
  \country{China}
}
\email{xingmz@zgclab.edu.cn}

\author{Haoyu Wang}
\orcid{0000-0003-1100-8633}
\affiliation{%
  \institution{Huazhong University of Science and Technology}
  \city{Wuhan}
  \country{China}
}
\email{haoyuwang@hust.edu.cn}

\author{Ying Gao}
\orcid{0000-0001-8992-651X}
\affiliation{%
  \institution{Beihang University}
  \city{Beijing}
  \country{China}
}
\affiliation{%
  \institution{Zhongguancun Laboratory}
  \city{Beijing}
  \country{China}
}
\email{gaoying@buaa.edu.cn}
\authornote{Ying Gao and Yinliang Yue are the corresponding authors.}

\author{Yinliang Yue}
\orcid{0000-0002-8417-2234}
\affiliation{%
  \institution{Zhongguancun Laboratory}
  \city{Beijing}
  \country{China}
}
\authornotemark[1]

\begin{abstract}
As Ethereum continues to thrive, the Ethereum Virtual Machine (EVM) has become the cornerstone powering tens of millions of active smart contracts.
Intuitively, security issues in EVMs could lead to inconsistent behaviors among smart contracts or even denial-of-service of the entire blockchain network.
However, to the best of our knowledge, only a limited number of studies focus on the security of EVMs. 
Moreover, they suffer from 1) insufficient test input diversity and invalid semantics; and 2) the inability to automatically identify bugs and locate root causes.
To bridge this gap, we propose {\framework}, a differential testing framework for EVM, which takes advantage of LLMs and static analysis methods to address the above two limitations.
We conducted the largest-scale evaluation, covering nine EVMs and uncovering 26 previously unknown bugs, 22 of which have been confirmed by developers and three have been assigned CNVD IDs.
Compared to state-of-the-art baselines, {\framework} can improve code coverage by at most 71.06\%, 148.40\% and 655.56\%, respectively. 
Through an analysis of real-world deployed Ethereum contracts, we estimate that 7.21\% of the contracts could trigger our identified EVM bugs under certain environmental settings, potentially resulting in severe negative impact on the Ethereum ecosystem.
\end{abstract}




\maketitle

\section{Introduction}
Ethereum is the second-largest blockchain platform, supporting smart contracts as its killer application~\cite{buterin2013ethereum}. Numerous applications are built upon smart contracts, such as NFTs~\cite{nft}, decentralized finance~\cite{defi}, and games~\cite{game}.
At the time of writing, the total market capitalization of Ethereum has reached up to \$307.21 billion ~\cite{CoinMarket}.
To support the execution of smart contracts, the Ethereum Virtual Machine (EVM) serves a critical role as the core execution environment.
Instead of running on each client node, EVMs can also be executed in local environments, such as simulating execution results for smart contracts during prototype testing~\cite{shou2023ityfuzz}.

Given that vulnerabilities in smart contracts can result in substantial financial losses, the security of smart contracts has drawn considerable attention from both academia and industry \cite{zhong2024prettysmart,zhang2024nyx,wu2024we,shou2023ityfuzz}.
In contrast, the security of the underlying EVM has been underexplored, which can result in even more severe negative impacts.
CVE-2021-39137~\cite{cve-2021-39137}, a consensus vulnerability in Go-Ethereum (Geth)~\cite{go-ethereum}, triggered a chain fork on Ethereum mainnet in August 2021. 
Due to Geth's widespread deployment as the dominant Ethereum client, this vulnerability allowed malicious contracts to fork Geth nodes from the mainnet as a result of inconsistent contract results, introducing double-spending risks.
The vulnerability stemmed from memory corruption in the EVM precompiled contract \texttt{datacopy}, highlighting the necessity for rigorous EVM implementation testing.

To the best of our knowledge, only a few tools are available for testing the EVM, including EVMFuzzer \cite{fu2019evmfuzzer}, NeoDiff \cite{maier2021NeoDiff} and go evmlab \cite{goevmlab, fuzzyvm}.
Specifically,
EVMFuzzer is the first differential testing tool for EVM, using  mutation on the source code of smart contracts to generate test inputs, while NeoDiff directly generates contract bytecode through predefined template.
Meanwhile, go evmlab employs state transition tests based on similar pre-defined templates.

However, testing EVMs are hindered by some inherent limitations.
On the one hand, \textit{it is challenging to generate diverse and semantically-valid test inputs}. Current work either generates test inputs on source-code-level through compilation, which greatly reduces diversity~\cite{fu2019evmfuzzer}, or only performs syntax-correct mutations on bytecode-level without considering semantic validity~\cite{maier2021NeoDiff,goevmlab,fuzzyvm}. It is hard to find a balance between these two requirements.
On the other hand, \textit{it is challenging to achieve automated bug identification and root cause localization for identified EVM inconsistencies.} Current approaches directly provide EVM developers with contracts that trigger inconsistencies, which brings a huge time and manpower burden to their subsequent debugging work.

\textbf{This work.}
We present {\framework}, an opcode-level differential testing framework for EVM, which can \textit{generate semantically-valid and diverse test inputs} for differential testing and \textit{automatically identify EVM implementation bugs with the corresponding root causes inside EVM}. 
We take advantage of Large Language Models (LLMs) with static analysis methods to extract opcode definitions from the specification, applying control-flow-oriented mutation and argument-oriented mutation to generate opcode-level test inputs. 
With collected results from differential testing, we focus on three critical metrics to identify bugs and propose a root cause localization algorithm leveraging LLMs to link the bug back to its corresponding implementation within the EVM.
We have evaluated the effectiveness of {\framework} on nine EVM implementations covering multiple scenarios.
Compared with state-of-the-art baselines, {\framework} increases code coverage by 71.06\%, 148.40\% and 655.56\%, respectively. Moreover, {\framework} has successfully uncovered 26 distinct bugs, 22 of which have been confirmed, and three of which are assigned CNVD IDs.

\textbf{Contribution.} We make the following contributions in this work:

\begin{itemize}
\item We present {\framework}, an opcode-level differential testing framework for EVM, which can generate semantical-valid and diverse test inputs by taking advantage of LLMs and static analysis.  

\item We propose an automated bug identification and root cause localization algorithm, leveraging LLMs to locate the faulty implementations in EVMs to facilitate risk assessment and bug patch.

\item {\framework} has identified 26 unknown bugs from nine EVM implementations, among which 22 have been confirmed or patched with our timely disclosure. Additionally, three vulnerabilities have been assigned with CNVD IDs.

\item Our evaluation of real-world Ethereum contracts indicates that opcodes with faulty implementations are widespread (7.21\% of existing contracts) within Ethereum mainnet.

\end{itemize}

\section{Background}

\subsection{Ethereum Virtual Machine}
\label{sec:backgroud:evm}

\textbf{Ethereum Virtual Machine.} The Ethereum Virtual Machine (EVM) serves as the runtime environment for executing Ethereum smart contracts. Designed as a stack-based virtual machine, the EVM executes bytecode at a low level. The architecture of the EVM encompasses several foundational concepts that are essential to Ethereum’s execution\cite{yellowPaper}:
\begin{itemize}
\item \textit{Stack.} The EVM operates with a stack-based architecture, maintaining a stack of up to 1024 elements, each represented as a 256-bit word.
\item \textit{Memory.} The memory model of EVM is a volatile, word-addressable byte array, with all memory locations initially defined as zero.
\item \textit{Gas.} To prevent network abuse and address challenges inherent to Turing completeness, all computations in the EVM incurs fees, measured in units of gas. The gas fees are charged on opcode level, whose price may fluctuate according to the concrete behavior of the opcode.  For instance, the EVM opcode \texttt{BALANCE} has a variable gas cost: if the address queried by \texttt{BALANCE} has already been accessed within the same transaction, the gas cost is 100; otherwise, a higher dynamic gas charge of 2600 applies.
\item \textit{Storage.} The storage in EVM is a persistent key-value store which is part of the Ethereum state, with all locations initially set to zero. Storage can be read and written using the specific opcodes, like \texttt{SLOAD} and \texttt{SSTORE}.
\item \textit{Code.} The bytecode of the smart contract to be executed is stored in an immutable ROM within EVM, accessible only through specialized opcodes (\textit{e.g.,} \texttt{CODESIZE}, \texttt{CODECOPY}).
\item \textit{Program Counter.} The program counter (pc) in the EVM indicates the position of the next opcode to be executed. For opcodes that include operands (\textit{e.g.,} \texttt{PUSH1 0x60}), the $pc$ is incremented by $1+x$, where $x$ represents the number of operands.
\end{itemize}

\noindent
\textbf{Ethereum Virtual Machine Opcodes.} \label{EVMbytecode}
An EVM opcode is an atomic instruction that the EVM can interpret and execute. Opcodes are building blocks of smart contract bytecode and are used to perform various operations within the EVM. Each opcode is represented by a single byte (8 bits), which allows for a range of 256 possible opcodes (from \texttt{0x00} to \texttt{0xFF} in hexadecimal). As of October 2024, Ethereum had undergone its latest mainnet upgrade, known as Dencun \cite{fork}, which has expanded the EVM to include 149 opcodes. 

\noindent
\textbf{Ethereum Specifications.} 
The Ethereum network is a continuously evolving decentralized system. To effectively test the security of the EVM, it is essential to understand its design and security features based on Ethereum specifications. Currently, the primary specifications for Ethereum include \textit{The Yellow Paper}, \textit{Ethereum Improvement Proposals (EIPs)}, and \textit{Ethereum Execution Client Specifications}.

\begin{itemize}
\item \textit{The Yellow Paper.} The Ethereum Yellow Paper \cite{yellowPaper} provides a mathematically rigorous formal specification of the Ethereum protocol and the EVM execution model. While its mathematical formalism ensures precision, it presents comprehensive challenges for programmers and has not been updated beyond the Shanghai fork of April 2023 \cite{fork}.
\item \textit{Ethereum Improvement Proposals.} Ethereum Improvement Proposals (EIPs) \cite{eips} are standards that define potential new features or processes for the Ethereum network. Once an EIP is accepted and implemented, it may modify or extend the specifications presented in the Yellow Paper, ensuring that the formal documentation remains consistent with protocol developments.
\item \textit{Ethereum Execution Layer Specifications.} For the purpose of providing a more programmer friendly and up-to-date specification of Ethereum than the traditional Yellow paper \cite{eelsblog}, the Ethereum Execution Layer Specification (EELS) \cite{exec-spec} was introduced. The EELS serves not only as an EVM implementation, but also as a specification of the core components of an Ethereum execution client. It is written in Python, with a focus on clarity and readability. EELS provides a current and detailed description of the Ethereum execution client specifications, integrating the latest protocol updates to ensure continued compatibility with the Ethereum mainnet.
\end{itemize}

\subsection{Differential Testing}
\label{sec:backgroud:testing}

Over the past decade, fuzzing has emerged as an exceptionally effective approach for uncovering software bugs \cite{fuzz-sok}. Fuzzing generally requires a test oracle—a mechanism to identify whether the output is correct or erroneous. In cases where an oracle is not easily available, differential testing comes to use.
Specifically, differential testing is used to identify bugs or inconsistencies by comparing the behavior or outputs of multiple implementations of the same functionality. In this approach, the same input is provided to each implementation, and their outputs are compared. Any discrepancies between the outputs can indicate unexpected behavior or potential bugs in one or more of the implementations. 
Differential testing first requires generating test inputs for target programs to execute. 
During the execution, the runtime information will be collected for comparison and further bug report.  Differential testing is particularly effective for uncovering semantic or logical bugs that do not exhibit explicit erroneous behaviors, such as crashes or assertion failures.

\section{Challenges \& Solution}
In this section, we will detail the challenges in performing differential testing for EVM by providing motivating examples in \S\ref{sec:challenge:motivation:example}. Then, we will propose our high-level solution in \S\ref{section:solution}.

\subsection{Motivating Example \& Challenges}
\label{sec:challenge:motivation:example}

Given the complexity of the EVM, fully avoiding implementation bugs remains challenging. To illustrate this, we present the following motivating example.
Fig.~\ref{fig:case:seal:fix} shows a bug (as well as its patch) in SealEVM \cite{sealevm}, a custom third-party EVM implementation.
Specifically, this bug is a mathematical calculation related bug within the implementation of opcode \texttt{EXP}. When executing the opcode \texttt{EXP} with crafted parameters, SealEVM will be stuck, disrupting the normal execution of subsequent opcodes.
The root cause is that SealEVM lacks the modulus implementation defined in the specification, as shown at Line 9 of Fig.~\ref{fig:exp:spec}.
Triggering this bug requires 1) a deep understanding of the semantics of the \texttt{EXP} opcode in the specification, and 2) careful construction of the corresponding parameters as input.

\begin{figure}[t]
    \centering
    \begin{subfigure}{\linewidth}
        \centering
\begin{lstlisting}[
            style=gocomment,
            numbersep=4pt,
            xleftmargin=1em,
            framexleftmargin=1em,
           ]
!!++ func expModulus() *big.Int {
++	     m := &big.Int{}
++	     return m.Exp(big.NewInt(2), big.NewInt(256), nil)
++ } !!

   func (i *Int) Exp(e *Int) *Int {
??--	     i.Int.Exp(i.Int, e.Int, nil) ??       
!!++	     i.Int.Exp(i.Int, e.Int, expModulus) !!
         return i.toI256()
   }
\end{lstlisting}
            \vspace{-0.1in}
        \caption{How SealEVM implements the \texttt{EXP} opcode, as well as the patch.}
        \Description{How SealEVM implements the \texttt{EXP} opcode, as well as the patch.}
        \label{fig:case:seal:fix}
    \end{subfigure}
    \begin{subfigure}{\linewidth}
        \centering
\begin{lstlisting}[
                     language=Python,    
                     numbers=left,    
                     numbersep=4pt,
                     xleftmargin=1em,
                    framexleftmargin=1em,
                     morekeywords={Evm, Uint, pop, charge_gas, GAS_EXPONENTIATION,GAS_EXPONENTIATION_PER_BYTE, GAS_HIGH, wrapping_add, push, InvalidJumpDestError}]
def exp(evm: Evm) -> None:
    base = Uint(pop(evm.stack))
    exponent = Uint(pop(evm.stack))
    exponent_bits = exponent.bit_length()
    exponent_bytes = (exponent_bits + Uint(7))
    charge_gas(
        evm, GAS_EXPONENTIATION + GAS_EXPONENTIATION_PER_BYTE * exponent_bytes
    )
    result = U256(pow(base, exponent, Uint(U256.MAX_VALUE) + Uint(1)))
    push(evm.stack, result)
    evm.pc += Uint(1)


\end{lstlisting}
                     \vspace{-0.1in}
        \caption{The specification of the \texttt{EXP} opcode.}
        \label{fig:exp:spec}
    \end{subfigure}
    \vspace{-0.3in}
    \caption{A motivating example about the bug of the \texttt{EXP} opcode in SealEVM.}
    \Description{A motivating example about the bug of the \texttt{EXP} opcode in SealEVM.}
    \vspace{-0.1in}
\end{figure}

\begin{figure}
    \centering
    \begin{subfigure}[t]{0.49\linewidth}
        \centering
\begin{lstlisting}[
                     mathescape=true,
                     language=Python,    
                     morekeywords={Evm, Uint, pop, charge_gas, GAS_VERY_LOW, GAS_HIGH, wrapping_add, push, InvalidJumpDestError}]
def jump(evm: Evm) -> None:
    jump_dest = Uint(pop(evm.stack))
    charge_gas(evm, GAS_MID)
    if jump_dest not in evm.valid_jump_destinations:
       raise InvalidJumpDestError
    evm.pc = Uint(jump_dest)
\end{lstlisting}
                     \vspace{-0.1in}
        \caption{The specification of the \texttt{JUMP} opcode.}
        \Description{The specification of the \texttt{JUMP} opcode.}
        \label{fig:jump:spec}
    \end{subfigure}
    \begin{subfigure}[t]{0.49\linewidth}
        \centering
\begin{lstlisting}[
                     language=go,    
                     morekeywords={PUSH1,JUMP}]
// bytecode 602056...6080
1  PUSH1 20 // Counter
3  JUMP     // Jump to counter

   ...

20 PUSH1 80 // Not a JUMPDEST 
\end{lstlisting}
                     \vspace{-0.1in}
        \caption{A semantically invalid case.}
        \Description{A semantically invalid case.}
        \label{fig:jump:example}
    \end{subfigure}
    \vspace{-0.1in}
    \caption{An example for illustrating a semantically-invalid case.}
    \vspace{-0.2in}
    \label{fig:motiv:example2}
\end{figure}

\begin{table}[t]
\caption{A comparison of existing differential testing tools for EVM, where \textbf{Level} and \textbf{Approach} refer to their test inputs generation methods. The \scalebox{0.7}{\faIcon{circle}}, \scalebox{0.7}{\faIcon{adjust}}, and \scalebox{0.7}{\faIcon[regular]{circle}} symbols represent full, partial, and no semantic validity or diversity, respectively.}
\vspace{-0.1in}
\label{table:challenges}
\resizebox{\textwidth}{!}{%
\begin{tabular}{c|cc|cc|cc}
\toprule
\multirow{2}{*}{\textbf{Tool}} &
  \multirow{2}{*}{\textbf{Level}} &
  \multirow{2}{*}{\textbf{Approach}} &
  \multicolumn{2}{c|}{\textbf{Challenge \#1}} &
  \multicolumn{2}{c}{\textbf{Challenge \#2}} \\
   &
   &
   &
  \textbf{Validity} &
  \textbf{Diversity} &
  \textbf{Bug Identification} &
  \textbf{Root Cause Localization} \\
  \midrule
EVMFuzzer & source code & compilation & \scalebox{0.7}{\faIcon{circle}} & \scalebox{0.7}{\faIcon[regular]{circle}}   & opcode length, gas & -      \\
NeoDiff   & bytecode    & template    & \scalebox{0.7}{\faIcon[regular]{circle}} & \scalebox{0.7}{\faIcon{adjust}}   & type-hash, state-hash & Opcode \\
FuzzyVM & bytecode    & template    & \scalebox{0.7}{\faIcon[regular]{circle}}  & \scalebox{0.7}{\faIcon{adjust}} & state transition & - \\    
\bottomrule 
\end{tabular}%
}
\vspace{-0.2in}
\end{table}

\label{sec:challenge}
In summary, conducting differential testing against different EVM implementations presents significant challenges, which can be summarized into two aspects in Table ~\ref{table:challenges}.

\noindent
\textbf{Challenge \#1: Generating test inputs that satisfy both diversity and semantic validity requirements is challenging.}
Testing the EVM implementation first requires the generation of test inputs. 
Currently, \textit{source-code-level} and \textit{bytecode-level} test input generation exist, which suffer from this challenge.

Specifically, source-code-level methods require source code as inputs to generate more test inputs. However, leveraging source code for testing has certain limits. 
According to the prior work of P. Ma \textit{et al.}~\cite{pmndss2024}, only 2\% of the contracts are open-source. In addition, our analysis of recently verified open source contracts (Etherscan dataset at \S\ref{sec:evaluation:bugs:char}) has shown that 88.42\% contracts are ERC20 contracts. 
Those ERC20 contracts represent real-world business logic, and attackers are unlikely to open source malicious contracts or share their bytecode.
This limits the characterization of malicious behaviors and hinders the effective testing of EVM implementations against corner cases.
In other words, \textit{the diversity of test inputs generated from source-code-level methods cannot satisfy the real-world requirements.}

As for bytecode-level methods, they directly take contract bytecode as inputs and perform mutation to enlarge the diversity of the test input corpus. However, the mutation only considers the syntax correctness without semantic constraints.
For example, Fig.~\ref{fig:jump:example} illustrates a \texttt{JUMP} related bytecode snippet. Though it meets the syntax correctness, like stack balance, it does not consider that \textit{each \texttt{JUMP} requires a corresponding \texttt{JUMPDEST}} as its semantic validity requirement, which is illustrated in Fig.~\ref{fig:jump:spec}.
Consequently, \textit{the bytecode-level methods would generate test inputs with semantic invalidity}.
 
\noindent
\textbf{Challenge \#2: It is challenging to achieve automated bug identification and root cause localization for EVM inconsistencies.}
Though differential testing does not require oracles, identifying the specific bug and localizing the corresponding root cause in EVM is still challenging.
Specifically, different EVM adopts different programming languages and component architectures in their implementations. Moreover, they may behave distinctively on the same metric. For example, due to language-specific features, EVM implementations handle errors differently across platforms. Consequently, \textit{identifying if bugs, \textit{i.e.,} discrepancies, really exist by comparing EVM behaviors is difficult.}
As for the root cause localization, it also puts forward further requirements on the scalability of the method. Specifically, only providing bytecode test input is of limited help to developers, because they do not know which opcode or which logic problem in the EVM causes this inconsistency. However, as mentioned above, different EVMs adopt different programming languages and architectural designs, leading to the challenges in locating the root cause.

\noindent
\textbf{Limitations of current tools.} To the best of our knowledge, only a few tools are available for testing the EVM, \textit{i.e.,} EVMFuzzer ~\cite{fu2019evmfuzzer}, NeoDiff ~\cite{maier2021NeoDiff}, and FuzzyVM ~\cite{fuzzyvm,goevmlab}.
We underline all existing tools have certain limitations. 

Specifically, EVMFuzzer is the first differential testing tool for EVM, utilizing source codes of smart contracts as seeds and employing 8 predefined mutators to generate new contracts. Subsequently, these contracts are compiled into bytecode for execution within the EVM. Intuitively, its method is limited by \textbf{Challenge \#1}. Moreover, it requires huge manual effort to confirm bugs and localize root causes, \textit{i.e.,} the ones stated in \textbf{Challenge \#2}.

As for NeoDiff, it is a feedback-guided differential fuzzing framework for EVMs.
It directly generates contract bytecode based on predefined templates and mutates them with feedback information. Due to the use of human-crafted templates, the generated bytecode exhibits similar structures, leading to insufficient diversity (\textbf{Challenge \#1}). Moreover, the excessive length of the NeoDiff generated bytecode complicates the process of bug identification and it is limited to reporting only the opcode responsible for the inconsistencies, lacking a comprehensive analysis of their underlying causes as discussed in \textbf{Challenge \#2}.  

FuzzyVM ~\cite{fuzzyvm}, maintained by developers from the Go-Ethereum Team ~\cite{go-ethereum}, generate test inputs based on pre-defined templates and performs fuzzing on Geth. The actual execution is handled by Go evmlab ~\cite{goevmlab}, which facilitates differential testing with other EVM implementations.
Specifically, FuzzyVM conducts state transition testing to find EVM bugs. State transition tests \cite{execution-spec-tests} first define the pre-state and execution environment of the EVMs, then execute a sequence of transactions to achieve a post-state. 
However, since the state test requires a well-defined execution environment setting and complicated transaction sequences, expert knowledge is necessary to write state tests effectively, which reflects the issues discussed in \textbf{Challenge \#1}. 
Besides, FuzzyVM requires manual effort to review inconsistencies, which also faces the \textbf{Challenge \#2}.

\subsection{Solution}
\label{section:solution}
In response to these two challenges, we propose the following ideas: 
To tackle \textbf{Challenge \#1}, we propose the concept of \textit{opcode-level test inputs generation}. It refers to generating short snippets for each opcode based on specifications.
To ensure the semantic validity and diversity of generated test inputs, we combine static analysis with Large Language Models (LLMs). Leveraging the advanced capabilities of LLMs in natural language understanding and the ICFG parsed by static analysis, as well as the \textit{control-flow-oriented} and \textit{argument-oriented} mutation strategies, we can generate semantically valid and diverse inputs tailored for differential testing.
Regarding \textbf{Challenge \#2}, we instrument the target EVMs and introduce three metrics designed to evaluate EVM implementation consistency for automated bug detection. For confirmed bugs, we propose a root cause localization algorithm leveraging LLMs to locate the responsible implementation in the source code. 

\section{Approach} 
This section presents the design of our framework, dubbed as {\framework}.

\subsection{Overview}
\begin{figure*}[t] 
    \centering 
    \includegraphics[width=1\columnwidth]{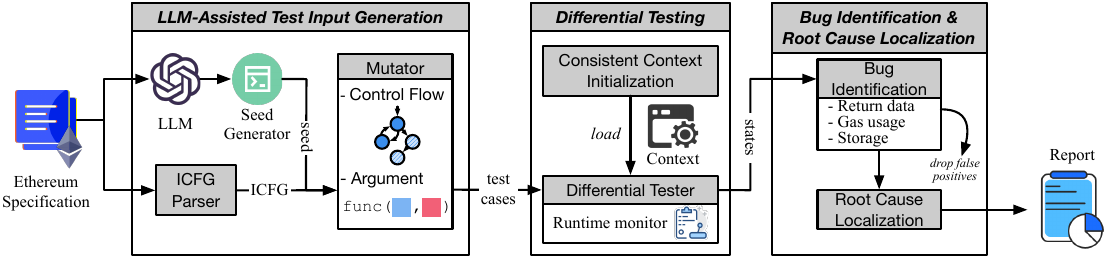} 
    \vspace{-0.3in}
    \caption{The workflow overview of \framework.} 
    \vspace{-0.1in}
    \label{fig:overview} 
    \Description{The overview of \framework.}
\end{figure*}

Fig.~\ref{fig:overview} illustrates the workflow of {\framework}.
As we can see, it can be divided into three phases, \textit{i.e.,} \textit{LLM-assisted test input generation} (\S\ref{section:approach:input:generation}), \textit{differential testing} (\S\ref{section:approach:execution}), and \textit{bug identification \& root cause localization} (\S\ref{section:approach:root:cause:analysis}). 
Specifically, {\framework} takes EELS (the specification for each opcode written in Python, see \S\ref{sec:backgroud:evm}) as inputs. Then, {\framework} will generate test inputs, while considering diversity and semantical validity, with the help of LLMs and static analysis.
Before conducting the differential testing, {\framework} will initiate the execution context for all EVMs.
After the execution, the results and runtime data from instrumented EVMs will be parsed in a uniform format for bug identification, which will be further utilized for root cause localization.
At last, {\framework} will output the corresponding bug reports.

\subsection{LLM-Assisted Test Input Generation}
\label{section:approach:input:generation}
Instead of using the existing source code of smart contract, which intertwines the implementation correctness of the compilers as described in \textbf{Challenge \#1}, we employ a generation-based method to directly generate EVM bytecode for testing. 
Initially, we employed template-based methods to generate test inputs derived from the Ethereum Execution Layer Specification (EELS). However, this approach requires significant human effort to read the specification and manually design the templates.
Considering the presence of code comments and definitions within specifications and the recent advancements demonstrated by Large Language Models (LLMs) in code generation and natural language processing tasks \cite{GPTScan,li2023nuances,liu2024make,xue2024llm4fin,guLLMBasedCodeGeneration2023}, LLMs offer an effective approach to interpret and generate test inputs based on opcode specifications.
In order to improve the performance of LLMs on this specific type of task, we take advantage of the EELS to construct prompts.
This phase can be further divided into four steps.

\begin{figure}[t]
    \prompt{Prompt 1: Opcode seed generator construction}{
    \textbf{System:} You are an expert in Ethereum Virtual Machine (EVM). \\
    \textbf{Task:} Given the following opcode specification, you should write a Python function that generates executable bytecode test case for the given opcode. \\
    \textbf{Requirements:}
    \begin{enumerate}[label=\arabic*.]
        \item The function should create a valid bytecode sequence, including any required push operations and memory operations for operands.
        \item The function should also add the bytecode corresponding to the opcode itself.
        \item Use random to generate random values where necessary, such as for operand lengths.
        \item The function should return the bytecode as a hex string.
    \end{enumerate}
    \textbf{Note:} There is no need to explain your answer, just return the Python code.
    
    \dashedline \\
    \textbf{Input:} EVM opcode \texttt{BYTE} and its specification.
\begin{lstlisting}[language=Python]
1   def get_byte(evm: Evm) -> None:
        """
        For a word (defined by next top element of the stack), retrieve the
        Nth byte (0-indexed and defined by top element of stack) from the
        left (most significant) to right (least significant).
        evm : The current EVM frame.
        """
        # STACK
2       byte_index = pop(evm.stack)
3       word = pop(evm.stack)
        # GAS
4       charge_gas(evm, GAS_VERY_LOW)
        # OPERATION
5       if byte_index >= 32:
6           result = U256(0)
7       else:
8           extra_bytes_to_right = 31 - byte_index
            # Remove the extra bytes in the right
9           word = word >> (extra_bytes_to_right * 8)
            # Remove the extra bytes in the left
10          word = word & 0xFF
11          result = U256(word)
12      push(evm.stack, result)
        # PROGRAM COUNTER
13      evm.pc += 1
\end{lstlisting}
    }
    \vspace{-0.15in}
    \caption{Prompt for generating seed generator of the \texttt{BYTE} opcode.}
    \vspace{-0.2in}
    \Description{Prompt for opcode \texttt{BYTE} seed generator construction.}
    \label{fig:template:generation}
\end{figure}

\subsubsection{Step I: Opcode Seeds Construction.}
\label{sec:approach:testcase:template}
To facilitate the test input generation, we leverage LLMs to build initial seeds for each opcode.
Specifically, we first explored the GitHub repository of EELS\footnote{EELS defines the EVM opcodes in the \texttt{src/ethereum/fork/vm/instructions} folder.} \cite{exec-spec}, where the semantics of each opcode is defined and written in Python functions.
For each opcode, we generate prompts to create the corresponding seed generators.
Fig.~\ref{fig:template:generation} is a prompt used for constructing a seed generator for the EVM opcode \texttt{BYTE}. As we can see, the prompt consists of context, task description, output requirements, additional notes, and the opcode definition from EELS.
Fig.~\ref{fig:byte:template} presents an example seed generator of the opcode \texttt{BYTE} opcode generated by LLMs.
The seed generator can generate a bytecode sequence, \textit{i.e., seed,} that correctly invokes the \texttt{BYTE} opcode by placing valid parameters on the stack, ensuring the semantical validity.

Existing research ~\cite{llmgencode} has demonstrated that GPT-4 could achieve high accuracy in Python code generation tasks, particularly for short code snippets. 
Our evaluation indicates that the length of the seed generator is an average of 14 lines of code.
Furthermore, our experimental results validate GPT-4's high accuracy, underscoring the feasibility of this approach.

\begin{figure}[t]
    \centering
\begin{lstlisting}[language=Python,numbers=none,morekeywords={randint, extend, generate_push_bytecode, fromhex}]
def generate_byte_bytecode() -> str:
    # Initialize an empty bytecode array
    bytecode = bytearray()
    # Generate and push a random value (1 to 32 bytes) onto the stack
    length1 = random.randint(1, 32)
    push_bytecode_hex1 = generate_push_bytecode(length1)
    push_bytecode1 = bytes.fromhex(push_bytecode_hex1)
    bytecode.extend(push_bytecode1)
    # Push a byte index (0x0 to 0xff) onto the stack
    length2 = 1
    push_bytecode_hex2 = generate_push_bytecode(length2)
    push_bytecode2 = bytes.fromhex(push_bytecode_hex2)
    bytecode.extend(push_bytecode2)
    # Append the BYTE opcode (0x1a)
    bytecode.extend(b"\x1a") 
    return bytecode.hex()
\end{lstlisting}
    \vspace{-0.15in}
    \caption{A generated seed generator for the \texttt{BYTE} opcode.}
    \vspace{-0.1in}
    \label{fig:byte:template}
    \Description{An Example Seed Generator for Opcode \texttt{BYTE}.}
\end{figure}

\begin{figure}[t]
    \centering
    \vspace{-0.05in}
    \begin{subfigure}[t]{0.35\linewidth}
        \centering
        \includegraphics[height=5cm]{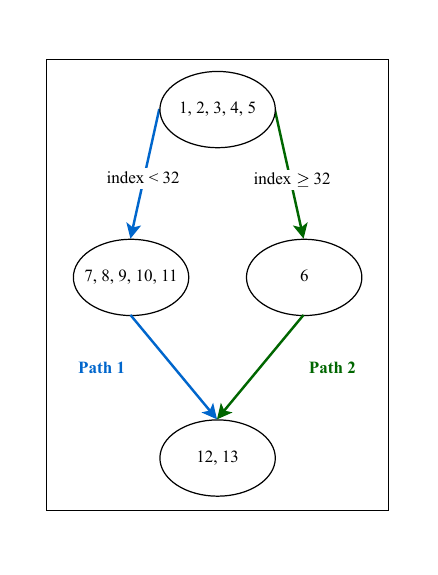}
        % \vspace{-0.1in}
        \caption{Without considering ICFG.}
        \label{fig:simple:cfg}
    \end{subfigure}
    \hfill
    \begin{subfigure}[t]{0.60\linewidth}
        \centering
        \includegraphics[height=5cm]{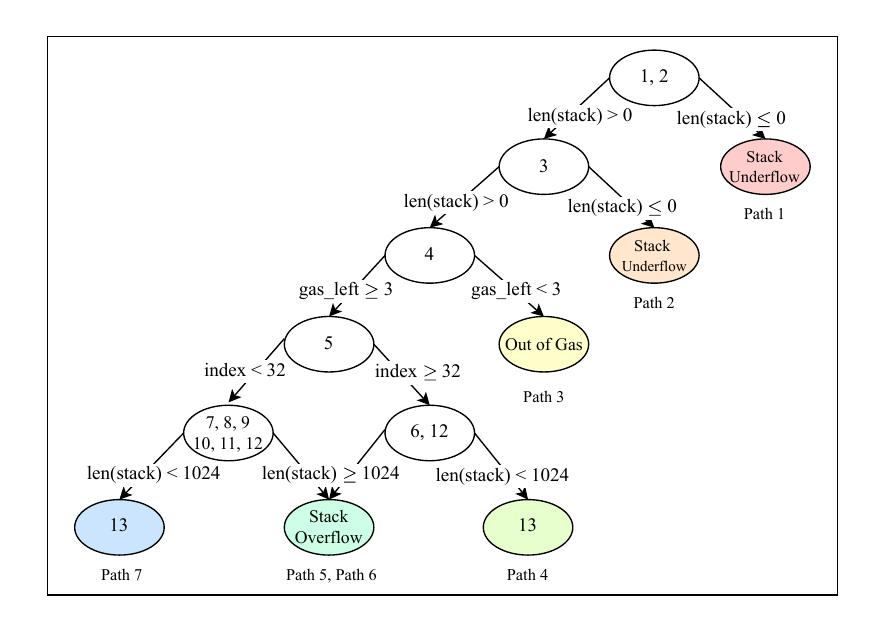}
        % \vspace{-0.1in}
        \caption{Considering ICFG.}
        \label{fig:full:cfg}
    \end{subfigure}
    \vspace{-0.1in}
    \caption{The control flow of \texttt{BYTE} in EVM, where the specification and line numbers in Fig.~\ref{fig:template:generation}.}
    \vspace{-0.2in}
    \label{fig:combined-figure}
    \Description{The specification of EVM opcode \texttt{BYTE} and its execution paths, derived from a single function and the complete control flow graph.}
\end{figure}

\subsubsection{Step II: Specification-Based ICFG Extraction.}
Though Step I illustrates that directly taking the EELS as input to construct seeds is feasible, since there may be function calls within the EELS (such as the \texttt{charge\_gas} call at Line 4 in Fig.~\ref{fig:template:generation}), considering only the specification function defined in EELS cannot cover its control flow completely.
For example, if external function calls are not taken into consideration, only two paths of opcode \texttt{BYTE} will be generated, as shown in Fig.~\ref{fig:simple:cfg}. However, callee may introduce additional control flows, if these are overlooked, it can prevent the generation of test inputs that cover the certain control paths each opcode may execute at runtime.
As shown in Fig.~\ref{fig:full:cfg}, the \texttt{pop} will introduce an additional control flow if the stack is empty, and the \texttt{charge\_gas} will revert if the given gas has been exhausted.
Therefore, we parse and extract the inter-procedural control flow graph (ICFG) for each opcode's specification, which could capture the control flow relationships across different functions. 

\subsubsection{Step III: Control-Flow-Oriented Mutation.}
% After we extracted opcodes inter-procedural CFG from the EELS, we collect semantic definitions of the EVM opcodes. 
To generate test inputs with diversity and semantical validity,  we utilize the reasoning and inferential capabilities of LLMs to mutate the existing seeds to cover more paths defined in the control flow of opcode specification. 
% Specifically, we construct prompts leveraging seeds and extracted ICFG to let LLM mutate the seed to other path in the ICFG.  
Specifically, to ensure that the test inputs generated by LLMs remain valid and executable, we leverage the previous seeds from \S\ref{sec:approach:testcase:template} to construct the input prompts. 
These seeds produced by seed generators serve as concrete examples that guide the LLMs in producing test inputs that adhere to the expected structure and specifications, ensuring semantical validity.
Considering that LLMs may struggle with interpreting hexadecimal bytecode seed, we convert the seed into its mnemonic representation to enhance the LLM's comprehension of the example. 
The input of the prompt consists of the opcode name, an example bytecode seed in both its hexadecimal and mnemonic forms, and an inter-procedural control flow graph (ICFG). 
Leveraging their reasoning and inferential capabilities regarding control flow within the ICFG, LLMs are expected to mutate the seeds and output a list of valid bytecode, tailored for execution within the EVM environment. An example prompt for control-flow-oriented mutation is shown in Fig.~\ref{fig:prompt:cfg}.

\begin{figure}
    \prompt{Prompt 2: Opcode control-flow-oriented mutation}{
    \textbf{System:} You are an expert in Ethereum Virtual Machine (EVM). \\
    \textbf{Task:} You will receive a JSON object with these four fields about an EVM opcode: \\
    \textbf{Requirements:}
    \begin{enumerate}[label=\arabic*.]
        \item opName: The name of the EVM opcode.
        \item seed: An example of a valid hex string bytecode seed  for the opcode \texttt{BYTE}.
        \item seed\_mnemonics: The mnemonic representation of the seed, which is helpful to understand the bytecode.
        \item ICFG: The inter-procedural control flow graph (ICFG) described in Python, covering all possible paths of the opcode's specification, constructed using the Graphviz dot tool. The bytecode provided in \texttt{seed} field represents one possible execution path in the ICFG.
    \end{enumerate}
    \textbf{Objective:}
    \begin{enumerate}[label=\arabic*.]
    \item Generate a comprehensive set of test inputs that cover all possible paths through the opcodes as defined by the provided ICFG.
    \item Each test input must be a valid hex string representing valid bytecode. You may refer to the format of the provided example, but you do not need to strictly follow it.
    \item Return the test inputs in a comma-separated list format, where each element is the corresponding hex string.
    \end{enumerate}
    \textbf{Input format:} A JSON object containing the four values above.\\
    \textbf{Output format:} Only a comma-separated list, where each element is a bytecode string executable within the EVM environment.\\
    \textbf{EXAMPLE OUTPUT:} \\
    \texttt{[hexstr\_1, hexstr\_2, ..., hexstr\_n]}  \\
    \textbf{Note:} No explanation is needed, just output the test inputs in a comma-separated list.  
    }
    \vspace{-0.15in}
    \caption{Prompt for implementing control-flow-oriented mutation for opcode \texttt{BYTE}.}
    \vspace{-0.2in}

    \Description{Prompt for implementing control-flow-oriented mutation for opcode \texttt{BYTE}.}
    \label{fig:prompt:cfg}
\end{figure}

\subsubsection{Step IV: Argument-Oriented Mutation.}
Considering that reaching certain branches in the control flow paths outlined in the opcode specification requires operands to be boundary values (For example, Path 4 in Fig.~\ref{fig:full:cfg}), LLMs may fail to mutate seeds to cover all the paths in the CFG.
Moreover, relying solely on LLMs to produce diverse bytecode is economically expensive. 
Therefore, we propose an argument-oriented mutation strategy to mutate the arguments in the mutated seed to introduce diversity and randomness for differential testing. 
The Algorithm \ref{algorithm:mutation} illustrates the details of argument-oriented mutation strategy. Specifically, as shown in Line 15, Line 17 in Algorithm \ref{algorithm:mutation}, we apply mutations to the operands following the \texttt{PUSHX} opcodes by replacing them with their boundary values or random values with a probability, where \texttt{X} is the byte length. 
This approach aims to introduce diversity in operand values, increasing the likelihood of reaching edge cases in the execution paths. 

\SetAlgoNoEnd
\SetAlgoSkip{}
\begin{algorithm}[t]
    \caption{The algorithm of argument-oriented mutation}
    \label{algorithm:mutation}
    \KwIn{$seed$ — the EVM bytecode seed to be mutated\\
    \hspace{2.8em} $p$ — the mutation probability\\
    \hspace{2.8em} $t$ — the number of mutation iterations}
    \KwOut{$testinputs$ — the set of test inputs after mutation}
    
    \SetKwFunction{FMutation}{Mutation}
    \SetKwProg{Fn}{Function}{:}{}
    \Fn{\FMutation{$seed$, $p$, $t$}}{
        $testinputs \gets \varnothing$\;
        \For{$iteration \gets 0$ \KwTo $t$}{
            $seed_{mut} \gets \texttt{""}$ \tcp*[r]{Initialize empty seed}
            \For{$index \gets 0$ \KwTo $seed.length$}{
                $opcode \gets seed[index:index+2]$\;
                $seed_{mut} \gets seed_{mut} \, || \, opcode$\;
                $index \gets index + 2$\;
                \If{$opcode \in \texttt{PUSHX\_SET}$}{
                    $byteslen \gets \texttt{getByteLen}(opcode)$\;
                    $index \gets index + 2 \times byteslen$\;
                    $r \gets \texttt{RandomFloat}(0,1)$\;
                    \If{$r < p$} { 
                        \If{$r < p/2$}{
                            $operand \gets \texttt{00} \times byteslen$ \tcp*[r]{All zeros}
                        }
                        \Else{
                            $operand \gets \texttt{FF} \times byteslen$ \tcp*[r]{All ones}
                        }
                    }
                    \Else{
                        $operand \gets \texttt{RandomBytes}(byteslen)$ \tcp*[r]{Random valid operand}
                    }
                    $seed_{mut} \gets seed_{mut} \, || \, operand$\;
                }
            }
            $testinputs \gets testinputs \cup seed_{mut}$ \tcp*[r]{Add the mutated seed to the set}
        }
        \Return{$testinputs$}
    }
\vspace{-0.04in}
\end{algorithm}

\subsection{Differential Testing}
\label{section:approach:execution}
In the second phase, {\framework} will conduct differential testing by taking the generated test inputs from \S\ref{section:approach:input:generation} as inputs on EVMs.
However, given the target EVMs are implemented in a different manner, we need to provide a consistent context for them. Moreover, instead of only examining  the final state for each EVM, we also analyze the intermediate states during the differential testing.

\subsubsection{Consistent Context Initialization.}
\label{section:consist:env}
To avoid false alarm during differential testing, we initialize all EVMs in a consistent context, including \textit{forks}, \textit{account states}, and \textit{global states}.

\begin{itemize}
    \item \textit{Forks.} Ethereum adopts \textit{forks} to introduce technical upgrades or changes \cite{fork}. In the context of EVM, different forks may represent different instruction sets, and even functionalities. To ensure the compatibility among EVMs, we configure all EVMs under the same fork, \textit{i.e., Cancun} ~\cite{fork}.
    \item \textit{Account states.} Ethereum can be taken as a state machine, where transactions can lead to the changes of permanent data, \textit{e.g.,} accout balance. To avoid the distinctions brought by these account states, we initialize a set of accounts under the same addresses with identical balances and embedded codes (\textit{i.e.,} smart contracts).
    \item \textit{Global states.} Apart from account states, Ethereum has several inherent global attributes, which may affect the state of the whole blockchain, \textit{e.g.,} chainid, block number, and timestamp. Similarly, to ensure the testing consistency, we must unify their values across EVMs.
\end{itemize}

For each round of differential testing, the context should be initialized consistently as we mentioned above. However, we have to argue that different context setting may lead to different behaviors. For example, the Path 3 in Fig.~\ref{fig:full:cfg} is related to the gas availability, which will be triggered exclusively by insufficient gas allocation.
Thus, before each round of testing, we will randomize the context while ensuring its consistency among EVMs.
Note that, since not all EVMs support custom context configurations, we had to modify the source code of certain EVM implementations to achieve such a consistency.

\subsubsection{Differential Testing.}
After setting the context, {\framework} takes test inputs generated from the ~\S\ref{section:approach:input:generation} as inputs and conducts differential testing.
Besides the final state, the runtime intermediate states, \textit{e.g.,} stack and memory, are also considered by {\framework} for the following bug identification and root cause localization.
Inspired by the EIP-3155 \cite{eip-3155}, we manually record the EVM state before the execution of each opcode in a same format.
Specifically, for each opcode, we pay attention to the program counter, the opcode name, remaining gas, consumed gas, stack, memory, and other auxiliary data.
This enables us to record necessary runtime information, ensuring comprehensive capture of execution details for each EVM.

\subsection{Bug Identification \& Root Cause Localization}
\label{section:approach:root:cause:analysis}
With collected states from the differential testing in the second phase, we specifically pay attention to some fields that are critical to cause EVM bugs. By leveraging LLMs, we can further pinpoint the corresponding functions in EVMs, reducing the burden of debugging for developers.

\subsubsection{Bug Identification.} 
\label{section:approach:root:cause:analysis:metric}

To identify bugs from EVM inconsistencies, we focus specifically on to the following three metrics abstracted from collected states in the second phase.
\begin{itemize}
\label{metrics}
    \item \textit{Return data.} The return data inside EVM is primarily utilized for messages calls to return values. It works similarly to a \textit{hash} representing the outcome of execution, directly reflecting inconsistency between EVMs. 
    \item \textit{Gas usage.} As the EVM is designed to use gas to restrict the resource abuse, EVM will charge the gas for each opcode before the operation is actually executed. If the gas usage is inconsistent, it means that either opcode is not implemented according to the specification, or a different control flow is taken by an EVM.
    \item \textit{Storage.} Storage is one of the permanent data stored on-chain. Intuitively, its changes can also be used as an indicator. In total, six opcodes are related to storage changes, \textit{i.e.,} \texttt{SSTORE}, \texttt{TSTORE}, \texttt{CREATE}, \texttt{CREATE2}, \texttt{SELFDESTRUCT}, and \texttt{REVERT}\footnote{Please refer to~\cite{evmcode-website} for their detailed explanations.}. 
    Before and after the execution of these six opcodes, we parse and record values in storage to identify if any inconsistency exists.
\end{itemize}

To ensure that the bug is reproducible, we will reproduce the identified bugs using the previously saved information. Specifically, we will re-execute the test input that triggered the bug on the EVMs with the same configuration as the prior run.
In summary, we will extract these three metrics from all intermediate and final states to assess any discrepancies between EVM implementations.
 
\SetAlgoSkip{}
\SetAlgoNoEnd
\begin{algorithm}[t]
\caption{Root Cause Localization Algorithm}
\label{algorithm:rootcause}
\KwIn{$testinput$: the EVM test input that triggered inconsistency\\
\hspace{2.8em} $sc$: the source code of the EVM}
\KwOut{$opcode$: the opcode responsible for the inconsistency\\
\hspace{3.5em} $\mathit{func}$: the function implementing the opcode\\
\hspace{3.5em} $cause$: the root cause} 
% \BlankLine
\SetKwFunction{FMain}{RootCauseLocalization}
\SetKwProg{Fn}{Function}{:}{}
\Fn{\FMain{$testinput, sc$}}{
    $ cause \gets \texttt{""}$\;
    $ runtime_{info} \gets \texttt{Execute}(testinput)$ \tcp*[r]{Get runtime information}
    $ opcode \gets \texttt{Compare}(\mathit{runtime_{info}})$ \tcp*[r]{Get the responsible opcode}
    % \tcp*[r]{Get the opcode responsible for the inconsistency}
    $ \mathit{funcMap} \gets \texttt{LLM4ExtractFunc}(opcode, sc)$ \tcp*[r]{Extract opcode-function map}
    \ForEach{$pair \in \mathit{funcMap}$}{
        \If{$ pair.key \texttt{ is } opcode$}{
            $ \mathit{func} \gets pair.value$\;
            $ \mathit{diff} \gets \texttt{getDiff}(\mathit{runtime\_info}, opcode)$ \tcp*[r]{Compare runtime information}
            \If{$\mathit{diff} \texttt{ found in } stack$}{
                $cause \gets cause \, || \, \text{"stack handling implementation"}$\;
            }
            \If{$\mathit{diff} \texttt{ found in } gas$}{
                $cause \gets cause \, || \, \text{"gas handling implementation"}$\;
            }
            \If{$\mathit{diff} \texttt{ found in } memory$}{
                $cause \gets cause \, || \, \text{"operation execution implementation"}$\;
            }
            \If{$\mathit{diff} \texttt{ found in } pc$}{
                $cause \gets cause \, || \, \text{"program counter handling implementation"}$\;
            }
        }
    }
    \Return{$(opcode, \mathit{func}, cause)$}
}
\vspace{-0.03in}
\end{algorithm}

\subsubsection{Root Cause Localization.}

After bugs are identified as mentioned in \S\ref{section:approach:root:cause:analysis:metric}, it is essential to locate their root causes to reduce duplicates and assist developers in fixing the corresponding issues.
Thus, we propose a \textit{root cause localization algorithm}, as shown in Algorithm~\ref{algorithm:rootcause}.
 
Firstly, we compare all values in fields of all states generated from different EVM implementations to locate the specific opcode that triggered the inconsistency (Lines 2 -- 4).
With the opcode corresponding to the identified bug, we further need to figure out which part of the implementation in EVM is responsible for that. 
We utilize LLMs to extract the mapping relations between opcodes and their corresponding functions in EVMs (Line 5).
Based on our analysis of the specification, we found that the logic in EELS for handling opcodes can be divided in to four phases: \textit{stack}, \textit{gas}, \textit{operation}, and \textit{pc}.
Therefore, when inconsistencies are detected, the root cause is identified by pinpointing the specific phase where the discrepancy occurs (Lines 10 -- 17).
For instance, if the divergence is observed in the stack values, it indicates the issue lies in the implementation on handling the EVM stack. 
Finally, the responsible opcode, its corresponding function in EVMs, as well as the root cause identified by Algorithm~\ref{algorithm:rootcause} are returned.

\section{Implementation and Evaluation}

We implemented {\framework} with Python, building on top of EVMFuzzer. 
To parse the ICFG from the Ethereum Execution Layer Specification, we use Scalpel \cite{li2022scalpel}, a static analysis framework for Python. 
For the integrated LLM, we choose the latest version of GPT-4o, the state-of-the-art LLM from OpenAI. Our experiments aim to answer the following four research questions: 

\begin{itemize}[leftmargin=2.5em]
    \item[\textbf{RQ1}] How is the performance of {\framework} compared to baselines?
    \item[\textbf{RQ2}] How many real-world bugs in EVM implementations can be identified by {\framework}, and what are their root causes?
    \item[\textbf{RQ3}] What are the characteristics and the real-world impacts of the detected bugs? 
    \item[\textbf{RQ4}] How does each component of {\framework} contribute to its code coverage?
\end{itemize}

\noindent \textbf{Experiment Setup \& Baseline Selection.} All experiments were performed on a desktop running Ubuntu 20.04 with an Intel i7-12700 CPU and 128GB RAM.
We selected the state-of-the-art EVM testing frameworks, EVMFuzzer \cite{fu2019evmfuzzer}, NeoDiff \cite{maier2021NeoDiff} and FuzzyVM \cite{fuzzyvm} as baselines for our evaluation.
For test input generation, EVMFuzzer operates at the source-code-level to produce bytecode for EVM execution, which faces challenges in \S\ref{sec:challenge}. In addition, NeoDiff generates bytecode-level test inputs using predefined templates, but similarly struggles with a lack of diversity, as discussed in \S\ref{sec:challenge}. FuzzyVM generates state test inputs using different strategies, which are also limited in diversity, as illustrated in \S\ref{sec:challenge}.

\begin{table}[t]
\caption{Overview of the selected EVM implementations.}
\vspace{-0.1in}
\label{target:evms}
\resizebox{0.7\textwidth}{!}{%
\begin{tabular}{clccc}
\toprule
% \textbf{EVM Project}  & \textbf{Category} & \textbf{Language} & \textbf{Github Stars} & \textbf{Version} \\
\textbf{Category} & \textbf{EVM Project} & \textbf{Language} & \textbf{Github Stars} & \textbf{Version} \\ \midrule
\multirow{3}{*}{execution client} & Geth ~\cite{go-ethereum}          & Golang     & 47.4k & 1.14.9  \\
                                  & Besu ~\cite{besu}          & Java       & 1.5k  & 24.8.0  \\
                                  & Nethermind ~\cite{nethermind} & C\#        & 1.3k  & 1.29.1  \\
\midrule
\multirow{4}{*}{standalone EVM}   & Py-EVM ~\cite{Py-EVM}        & Python     & 2.3k  & 0.8.0b1 \\
                                  & evmone ~\cite{evmone}        & C++        & 867   & 0.11.0  \\
                                  & ethereumjs ~\cite{ethereumjs/evm} & TypeScript & 2.6k  & 2.1.0   \\
                                  & revm  ~\cite{revm}         & Rust       & 1.6k  & 0.9.0   \\
\midrule
\multirow{2}{*}{custom EVM}       & SealEVM  ~\cite{sealevm}      & Golang     & 16    & 0.3.0   \\
                                  & Chainmaker ~\cite{chainmaker}    & Golang     & 3     & 2.3.3   \\ 
\bottomrule
\end{tabular}
}
\vspace{-0.1in}
\end{table}

\label{sec:eval:target:evms}
\noindent \textbf{Targeted EVMs.}
To systematically test the EVM implementations, we collected 49 EVM implementations in the Ethereum ecosystem to the best of our efforts.
Out of them, we choose three categories of EVM as our targets, which together cover more than 97\% execution clients on the Ethereum mainnet \cite{diversity} and include all officially recommended standalone EVM implementations \cite{standalone}. Besides, to test the completeness of minority third-party EVM implementations, we also select  two relatively niche EVMs for analysis. 
Our target EVMs can be divided into three categories, \textit{i.e.,} \textit{Ethereum mainnet execution client's EVM}, \textit{Ethereum officially recommended standalone EVM}, and \textit{Third-party custom EVM}, as shown in Table \ref{target:evms}.
All these EVMs are actively maintained by developers, with their most recent commits pushed within the past two months.

\begin{table}[t]
\caption{Code coverage results for 12 hours.}
\vspace{-0.1in}
\label{table:ablation}
\resizebox{0.9\textwidth}{!}{%
\begin{tabular}{c|cc|cc}
\toprule
\textbf{Tool} & \textbf{Geth (All)} & \textbf{Geth (Opcode Funcs)} & \textbf{SealEVM (All)} & \textbf{SealEVM (Opcode Funcs)} \\
\midrule
\rowcolor{codegray} 
{\framework}  & 54.40\% & 91.85\%   & 90.10\%  & 89.87\%      \\ 
EVMFuzzer & 31.80\% & 44.12\%   & 56.60\%  & 57.38\%      \\
NeoDiff   & 21.90\% & 17.98\%   & N/A  & N/A           \\
FuzzyVM   & 7.20\%  &  1.43\%   & N/A  & N/A           \\
\bottomrule
\end{tabular}%
}
\vspace{-0.2in}
\end{table}

\subsection{RQ1: Comparison with Baselines}
\begin{figure}[t]
    \centering
    \begin{subfigure}{0.49\linewidth}
        \includegraphics[width=1\textwidth]{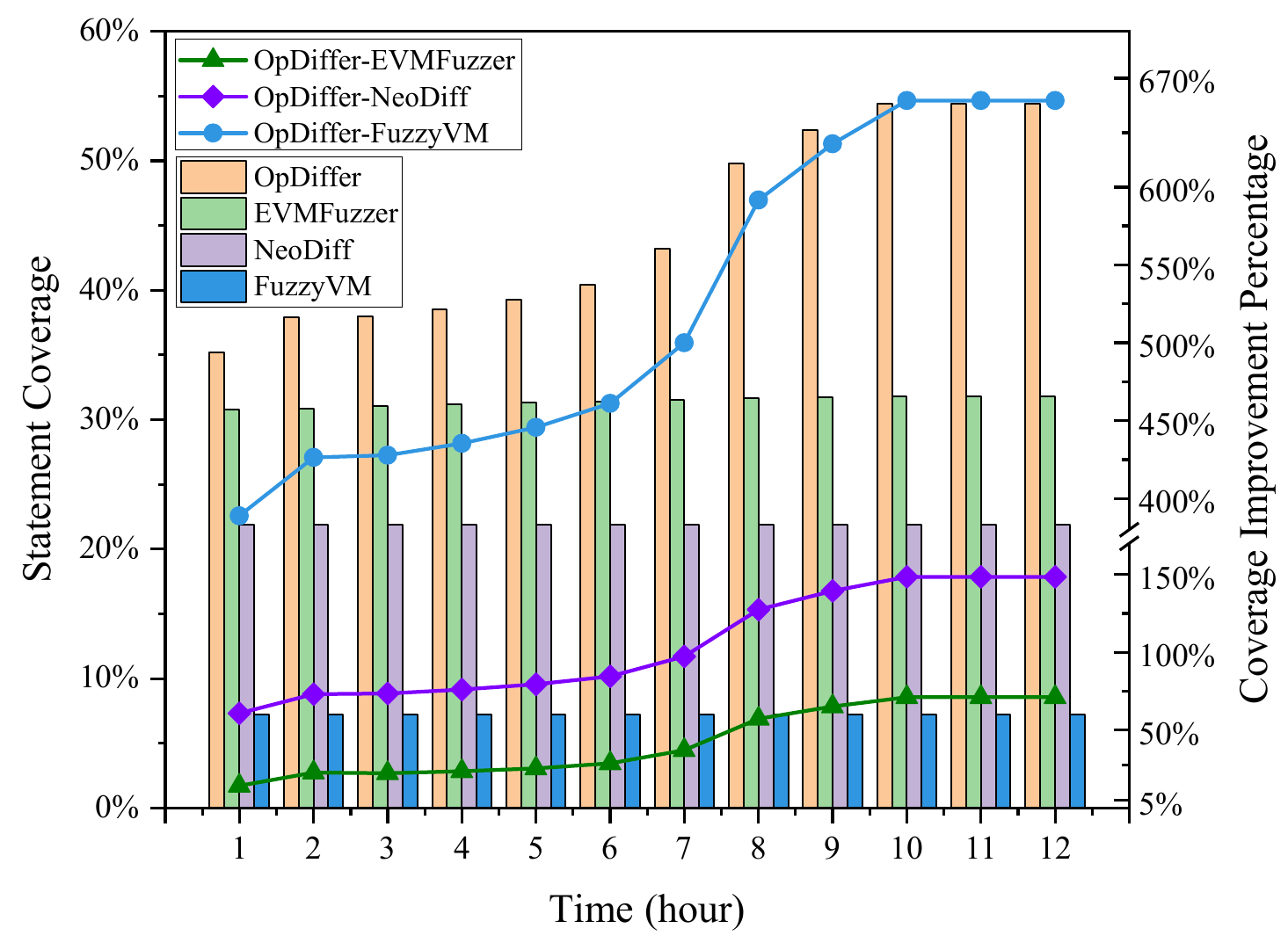}
        \caption{Geth (All)}
        \label{fig:geth:vm:covergae}
    \end{subfigure}
    \begin{subfigure}{0.49\linewidth}
        \includegraphics[width=1\textwidth]{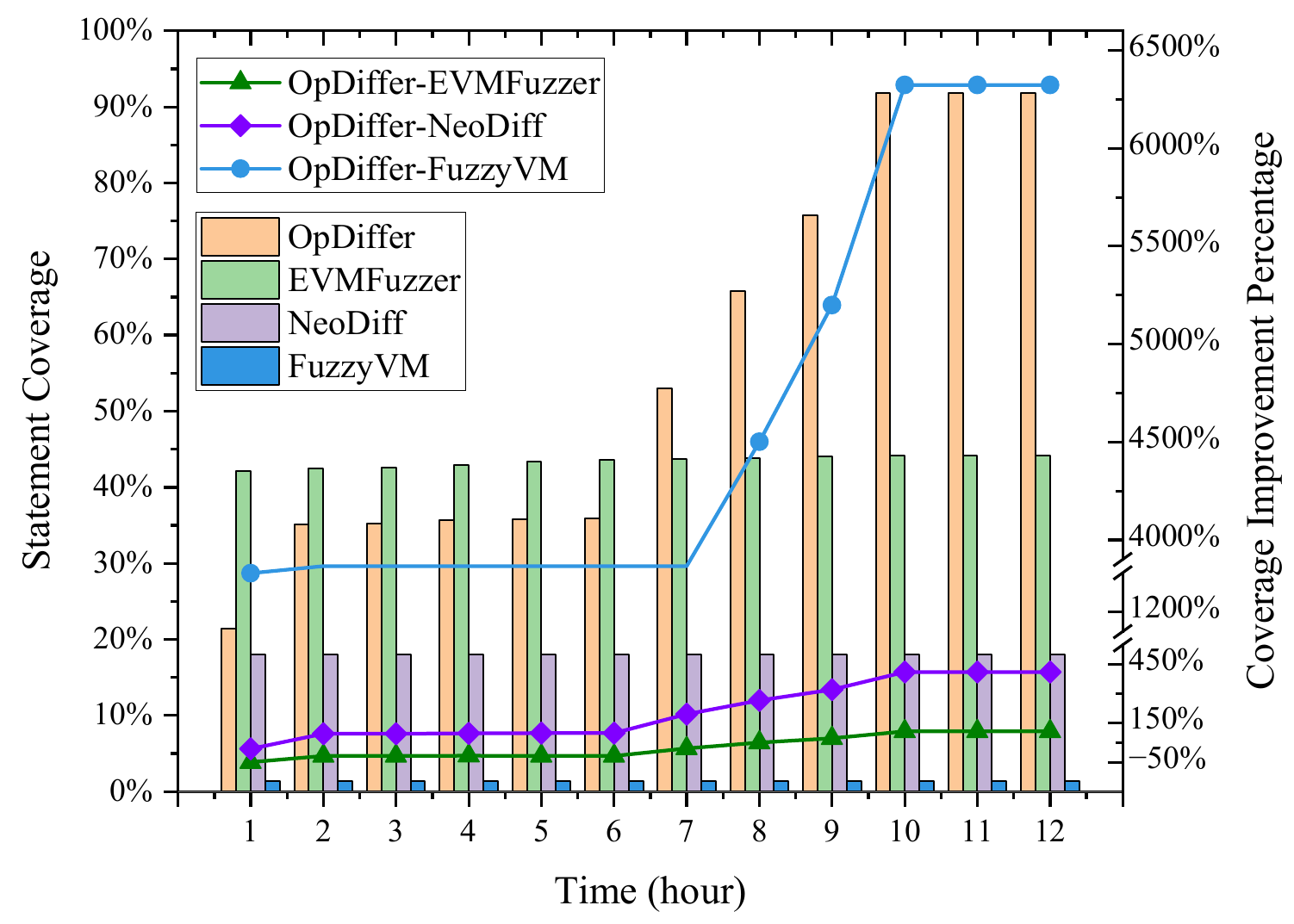}
        \caption{Geth (Opcode Funcs)}
        \label{fig:geth:func:covergae}
    \end{subfigure}
    \begin{subfigure}{0.49\linewidth}
        \includegraphics[width=1\textwidth]{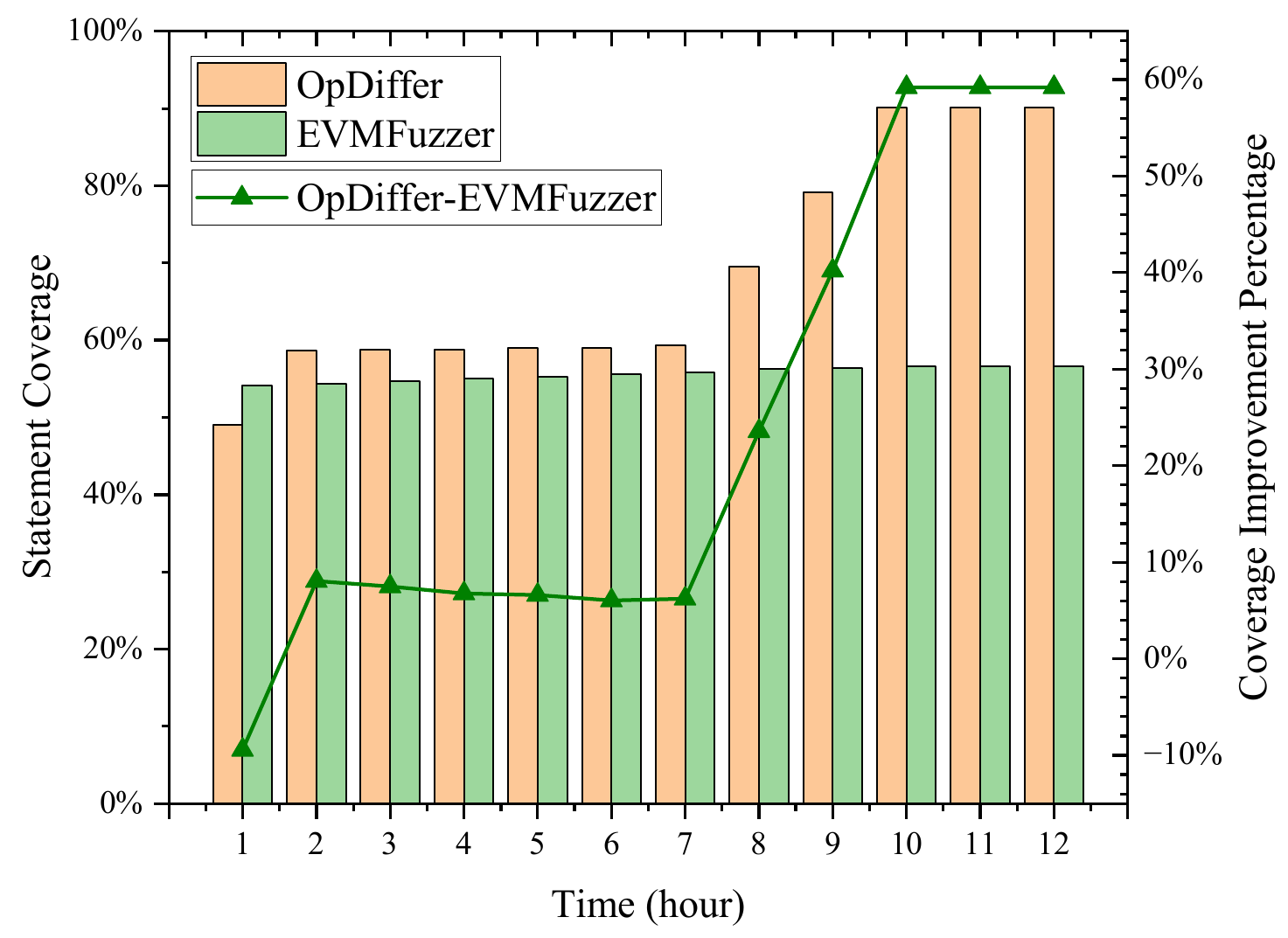}
        \caption{SealEVM (All)}
        \label{fig:seal:vm:covergae}
    \end{subfigure}
    \hfill
    \begin{subfigure}{0.48\linewidth}
        \includegraphics[width=0.99\textwidth]{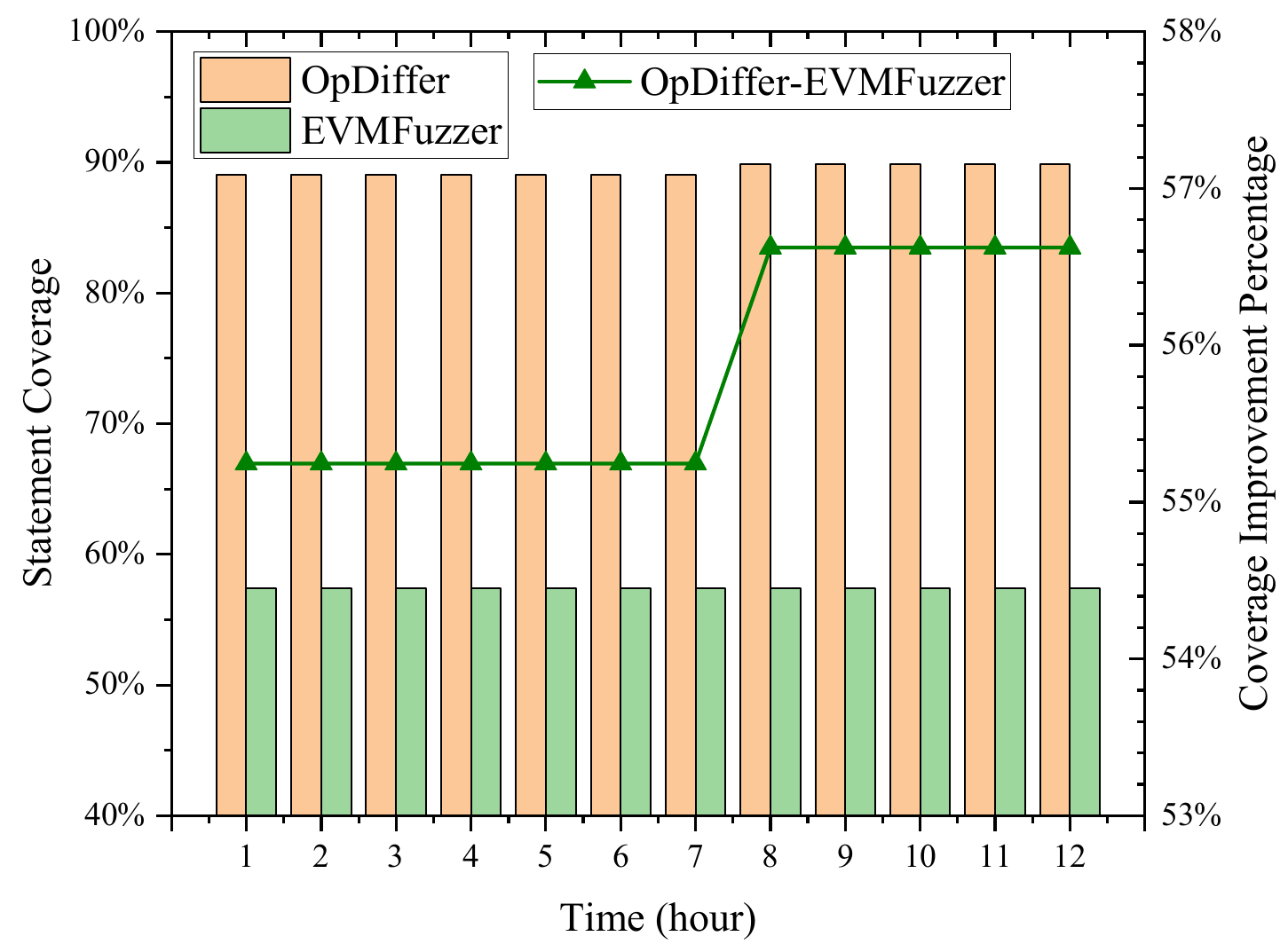}
        \caption{SealEVM (Opcode Funcs)}
        \label{fig:seal:func:covergae}
    \end{subfigure}
    \vspace{-0.1in}
    \caption{Coverage results of {\framework}, EVMFuzzer, NeoDiff and FuzzyVM on Geth and SealEVM. \textbf{Bars} represent the achieved code coverage, while \textbf{lines} represent the coverage improvement by which {\framework} exceeds EVMFuzzer, NeoDiff and FuzzyVM.}
    \label{fig:coverage}
    \Description{Coverage results of {\framework}, EVMFuzzer, NeoDiff and FuzzyVM on Geth and SealEVM. \textbf{Bars} represent the achieved code coverage, while \textbf{lines} represent the coverage improvement by which {\framework} exceeds EVMFuzzer, NeoDiff and FuzzyVM.}
\end{figure}

To compare the performance of {\framework} with baselines, we mainly focus on their code coverage on EVMs.
As Geth is the only EVM supported by all tools, we first take Geth into consideration. As {\framework} is implemented based on EVMFuzzer, thus both of them support SealEVM. Finally, against Geth and SealEVM, we ran {\framework}, EVMFuzzer, NeoDiff and FuzzyVM for 12 hours.
As both targeted EVMs are implemented in Golang, we take advantage of its official code coverage tool \cite{gocov}.
We measured not only the code coverage of the EVM as a whole, but also the specific functions responsible for opcode processing.
We recorded the code coverage on an hourly basis, where the results are shown in Fig.~\ref{fig:coverage} and Table \ref{table:coverage}.

\noindent \textbf{Overall Results.}
As shown in Table \ref{table:coverage}, when considering Geth/evm as a whole, {\framework} achieves a code coverage of 54.4\%, surpassing the coverage of EVMFuzzer, NeoDiff, and FuzzyVM by 71.06\%, 148.40\% and 655.56\%, respectively.
When only considering the opcode-related functions, {\framework} can cover 91.85\% of statements, which is 108.18\%, 410.85\% and 6323.08\% more than EVMFuzzer, NeoDiff and FuzzyVM, respectively.
FuzzyVM reached the lowest coverage for Geth that only 1.43\% of the lines in the opcode implementation functions were touched. This may be attributed to its inability to generate semantically valid and diverse test inputs.
In the case of SealEVM, the results are similar.
Compared to EVMFuzzer, {\framework} achieves 59.18\% and 56.62\% higher code coverage when considering SealEVM as a whole or only opcode-related functions.
The reason behind such a significant improvement is that {\framework} proposes an opcode-level test input generation approach based on the EVM specification, which can generate semantically valid and diverse test inputs with LLM's assistance. In contrast, due to EVMFuzzer generates bytecode through compilation, while NeoDiff and FuzzyVM utilizes templates-based approach for test input generation; these tools struggle to produce semantic-valid and diverse test inputs capable of covering certain statements in EVM implementations.

\noindent \textbf{Coverage Trend.}
We measured the code coverage on an hourly basis. After twelve hours, the coverage results of all tools eventually converge, which are displayed in Fig.~\ref{fig:coverage}.

As shown in Fig.~\ref{fig:geth:vm:covergae}, {\framework}  achieves higher coverage than EVMFuzzer, NeoDiff and FuzzyVM throughout the entire testing process. 
The coverage of other tools has converged within less than one hour, indicating that existing tools fail to test the Geth implementations of EVM opcodes due to lack of diversity of test inputs. 
Although baselines may generate more test inputs than {\framework}, their generated inputs cannot further improve the coverage across target EVM implementations.
In contrast,  {\framework} generates more diverse and semantically valid test inputs, resulting in its increase in coverage even after eight hours. 
In addition, Fig.\ref{fig:geth:func:covergae} demonstrates {\framework}'s coverage of opcode-related functions compared to baselines. While EVMFuzzer initially shows higher coverage by testing multiple opcodes simultaneously through compiler-generated bytecode, {\framework}'s depth-first strategy focuses on thorough testing of individual opcodes. Despite a lower initial coverage in the first five hours, {\framework} ultimately achieves 59.18\% and 56.62\% higher coverage than EVMFuzzer after 12 hours on SealEVM, as shown in Fig.\ref{fig:seal:vm:covergae} and Fig.~\ref{fig:seal:func:covergae}.

\answerbox{\textbf{\noindent Answer for RQ1: } {Compared with three state-of-the-art baselines, {\framework} achieves at most 148.40\% and 410.84\% coverage improvement when considering the EVM as a whole and only the opcode-related functions, respectively. The continuous growth of the code coverage proves the effectiveness of our proposed test input generation method.
}}

\begin{table}[h]
\caption{Details of all identified bugs of {\framework}, \textbf{Qty} is the number of affected opcodes, \textbf{Opcode} is an example affected opcode. Symbols \scalebox{0.7}{\faIcon{circle}},\scalebox{0.7}{\faIcon{adjust}}, and \scalebox{0.7}{\faIcon[regular]{circle}} represent bugs assigned CNVD/CVE IDs, currently under review, or not reported, respectively.}
\label{table:real:world:bugs:short}
\resizebox{0.9\textwidth}{!}{%
\begin{tabular}{cccccccc}
\toprule
\textbf{No.} & \textbf{Project} & \textbf{State} & \textbf{Root Cause} & \textbf{Opcode} & \textbf{Qty} & \textbf{Issue} & \textbf{CNVD} \\ \midrule
1  & \multirow{4}{*}{Geth}       & Fixed     & Network fork     & PUSH0        & 6 & \#29782 & \scalebox{0.7}{\faIcon[regular]{circle}} \\
2  &                             & Fixed     & Instruction set  & PREVRANDAO   & 1 & \#29722 &\scalebox{0.7}{\faIcon{adjust}}    \\
3  &                             & Fixed     & Address setting  & BALANCE      & 1 & \#30254 & \scalebox{0.7}{\faIcon[regular]{circle}}    \\
4  &                             & Fixed     & Fee setting      & BASEFEE      & 1 & \#30279 & \scalebox{0.7}{\faIcon[regular]{circle}}    \\ \midrule
5  & \multirow{4}{*}{Besu}       & Fixed     & Gaslimit setting & GASLIMIT     & 1 & \#7432  & \scalebox{0.7}{\faIcon[regular]{circle}}    \\
6  &                             & Fixed     & Fee setting      & BASEFEE      & 1 & \#7433  & \scalebox{0.7}{\faIcon[regular]{circle}}    \\
7  &                             & Fixed     & Tracing          & REVERT       & 1 & \#7608  & \scalebox{0.7}{\faIcon[regular]{circle}}    \\
8  &                             & Fixed     & Entry point      & N/A          & N/A & \#7430  & \scalebox{0.7}{\faIcon[regular]{circle}}    \\ \midrule
9  & Nethermind                  & Confirmed & Instruction set  & PUSH0        & 6 & \#7629  & \scalebox{0.7}{\faIcon[regular]{circle}}    \\ \midrule 
10 & \multirow{2}{*}{EthereumJS} & Confirmed & Address setting  & N/A            & N/A & \#3236  & \scalebox{0.7}{\faIcon{circle}}     \\
11 &                             & Will fix  & Memory tracing   & N/A            & N/A & \#3561  & \scalebox{0.7}{\faIcon[regular]{circle}}    \\ \midrule
12 & \multirow{2}{*}{Py-EVM}     & Fixed     & Missing field    & SELFDESTRUCT & 1 & \#2176  & \scalebox{0.7}{\faIcon{circle}}     \\
13 &                             & Reported  & Fee setting      & N/A            & N/A & \#2162  & \scalebox{0.7}{\faIcon{circle}}     \\ \midrule 
14 & \multirow{2}{*}{revm}       & Confirmed & Fee setting      & N/A            & N/A & \#1702  & \scalebox{0.7}{\faIcon[regular]{circle}}    \\
15 &                             & Reported  & Address setting  & N/A            & N/A & \#1704  & \scalebox{0.7}{\faIcon[regular]{circle}}    \\ \midrule
16 & \multirow{2}{*}{EELS}       & Reported  & Memory           & N/A            & N/A & \#981   & \scalebox{0.7}{\faIcon[regular]{circle}}    \\
17 &                             & Reported  & Description      & N/A            & N/A & \#992   & \scalebox{0.7}{\faIcon[regular]{circle}}    \\ \midrule
18 & \multirow{6}{*}{SealEVM}    & Will fix  & Fee setting      & N/A            & N/A & \#18    & \scalebox{0.7}{\faIcon[regular]{circle}}    \\
19 &                             & Fixed     & Stack            & CREATE2      & 1 & \#21    & \scalebox{0.7}{\faIcon[regular]{circle}}    \\
20 &                             & Fixed     & Pc manipulation  & N/A            & N/A & \#27    & \scalebox{0.7}{\faIcon{adjust}} \\
21 &                             & Fixed     & Memory           & MSIZE        & N/A & \#28    & \scalebox{0.7}{\faIcon[regular]{circle}}    \\
22 &                             & Fixed     & Storage          & EXTCODESIZE  & N/A & \#30    & \scalebox{0.7}{\faIcon[regular]{circle}}    \\
23 &                             & Fixed     & Math             & EXP          & 6 & \#31    & \scalebox{0.7}{\faIcon{adjust}} \\ \midrule
24 & \multirow{3}{*}{ChainMaker} & Fixed     & Stack            & CREATE2      & 1 & \#1092  & \scalebox{0.7}{\faIcon[regular]{circle}}    \\
25 &                             & Fixed     & Pc manipulation  & N/A            & N/A & \#1140  & \scalebox{0.7}{\faIcon{adjust}} \\
26 &                             & Fixed     & Math             & EXP          & 6 & \#1173  & \scalebox{0.7}{\faIcon{adjust}} \\ 
\bottomrule
\end{tabular}%
}
\vspace{-0.2in}
\end{table}

\subsection{RQ2: Real-World EVM Bugs}
\label{sec:eval:evmbugs}
We have shown that {\framework} can achieve higher coverage than the state-of-the-art baselines. We will further explore the capability of {\framework} in detecting real-world EVM bugs.

In total,  {\framework} has identified 26 bugs, whose detailed information is shown in Table \ref{table:real:world:bugs:short}.
To assist further investigation, we prepare detailed and reproducible steps and potential root causes for every identified bug.
We collect all identified bugs as well as the output reports and report them to the corresponding EVM developers.
As we can see, of the 26 identified bugs, 22 of them have been confirmed by the developers already, and 17 have been fixed at the time of writing.
Additionally, the developers have indicated plans to fix two of the remaining issues in future updates.  
We further conducted manual analysis to identify if there are security vulnerabilities.
Among the 26 confirmed bugs, three vulnerabilities have been assigned with China National Vulnerability Database (CNVD) IDs, while five of them are still under review by CNVD and NVD.
From Table \ref{table:real:world:bugs:short}, we can observe that:
1) Regardless of which of the three categories the EVM falls into, bugs affect them uniformly.
2) Developers have an active attitude towards the bugs identified by {\framework}, as they promptly confirmed and fixed the bugs with our assistance. 
We attribute this responsiveness to the comprehensive and reproducible bug reports provided, which streamline the debugging and validation processes for developers.  
3) Execution client EVM implementations have the fewest security-related bugs due to extensive testing before their deployment.  In contrast, standalone and custom EVMs remain vulnerable to several bugs, underscoring the critical importance of rigorous testing across all EVM implementations.

\answerbox{\textbf{\noindent Answer for RQ2: }
Out of nine Ethereum mainnet execution client EVMs, standalone EVMs, and third-party custom EVMs, {\framework} can successfully identify 26 bugs. With our timely disclosure, 22 bugs have been confirmed, while 3 of them may lead to severe impacts and have been assigned CNVD IDs. }

\subsection{RQ3: Bug Characterization}
\label{sec:evaluation:bugs:char}

To further illustrate the impact of identified bugs on blockchain systems, we assess the real-world consequences of these EVM bugs by analyzing smart contracts from the Ethereum mainnet.
Additionally, we present case studies for two critical identified bugs.

\begin{table}[t]
\caption{Statistics of two datasets that are used to evaluate the real-world impact for identified buggy opcodes.}
\vspace{-0.1in}
\label{table:buggy:rates}
\resizebox{0.7\textwidth}{!}{%
\begin{tabular}{ccccc}
\toprule
\textbf{Dataset} & \textbf{\#Contracts} & \textbf{\#Opcodes} & \textbf{\#Buggy 
 Opcodes} & \textbf{Buggy Rate} \\
\midrule
Gigahorse & 10,000 & 55,777,530 & 1,874,626 & 3.36\% \\
Etherscan & 10,000 & 58,979,456 & 4,253,359 & 7.21\% \\
\bottomrule
\end{tabular}
}\vspace{-0.2in}
\end{table}

\noindent \textbf{Real-world impact.}
Out of 26 identified bugs, there are 21 opcodes that are responsible for them, accounting for 14.1\% of all valid opcodes defined in the specification.
To evaluate the real-world impact of our detected bugs, we assume that if a contract contains one of the buggy opcodes, when it is executed on the corresponding EVM with specific context, the bug can be triggered.
We have to argue that this rate is the upper bound of real-world impacts of detected EVM bugs as there is still much effort left to exploit the vulnerability. 
However, an attacker could deploy malicious contracts containing buggy opcodes on the mainnet and send specific transactions designed to trigger the associated bugs in the affected EVM implementations. Such an attack could disrupt normal clients by causing issues during block synchronization and the execution of transactions.

Specifically, we collected two datasets. 
One is from Gigahorse \cite{grech2019gigahorse}, consisting of 10,000 compiled contract bytecode, while another is 10,000 recently verified contracts from Etherscan \cite{etherscan}, as certain buggy opcodes are introduced in recent Ethereum forks, like \texttt{PUSH0} and \texttt{MCOPY}.
The second dataset was compiled using the latest version of the Solidity compiler (solc 0.8.28). Our analysis indicates that 88.42\% of the contracts in this dataset conform to the ERC20 standard, which reflects the homogeneity of the source-code-level test input generation methods.

As presented in Table \ref{table:buggy:rates}, our experimental results indicate that 1,874,626 and 4,253,359 opcodes respectively in the two datasets could be affected by our identified EVM bugs.
The corresponding rates of affected opcodes are 3.36\%  and 7.21\%, respectively.
This evaluation demonstrates that the opcodes with EVM implementation bugs we detected are widely distributed across the Ethereum ecosystem, highlighting their widespread existence in real-world smart contracts. Although the proportion of potentially affected opcodes is currently around 3.36\% on the Ethereum mainnet, this rate could increase when new compiler features are enabled and become more widely used.

\begin{figure}[t]
    \centering
\begin{lstlisting}[
        language=Python,numbers=left,    
        numbersep=4pt,
        xleftmargin=1em,
        framexleftmargin=1em,
        morekeywords={get_opcode_fn,trace,mnemonic}]
for opcode in computation.code:
    opcode_fn = computation.get_opcode_fn(computation.opcodes, opcode)
    computation.logger.trace(
        "OPCODE: 0x%x (%s) | pc :%s",
        opcode,
# Bug here: the mnemonic field for the SELFDESTRUCT func is not defined
        opcode_fn.mnemonic, 
        max(0, computation.code.pc - 1),
    )
    ...
\end{lstlisting}
        \vspace{-0.15in}
    \caption{Case study \#1: the buggy implementation of \texttt{SELFDESTRUCT} in Py-EVM.}
    \Description{Case study \#1: the buggy implementation of \texttt{SELFDESTRUCT} in Py-EVM.}
    \vspace{-0.15in}
    \label{fig:case:pyevm}
\end{figure}

\noindent \textbf{Case \#1.}
The first case is the 12th entry in Table \ref{table:real:world:bugs:short}, which has been assigned with CNVD-2024-40890.
This is a bug of Py-EVM 0.8.0b1, which is a Python implementation of EVM.
It is related to the Py-EVM implementation for opcode \texttt{SELFDESTRUCT}, as shown in Fig.~\ref{fig:case:pyevm}.
As we can see, before executing the opcode, Py-EVM will first look up the implementation function \texttt{opcode\_fn} for the specific opcode (Line 2).
However, because in the function for \texttt{SELFDESTRUCT}, the field \texttt{mnemonic} is not defined yet (Line 7).
As a result, Py-EVM raises an error and halts the execution for following opcodes. Attacker could exploit this vulnerability by crafting malicious bytecode, enabling a denial-of-service (DoS) attack on Py-EVM.

\noindent \textbf{Case \#2.}
This case includes the 2nd entry in Table \ref{table:real:world:bugs:short}, related to the well-known Geth.
According to statistics \cite{diversity}, Go-Ethereum (Geth) is the most widely used execution client on the Ethereum mainnet as of October 2024.
We find that when executing the opcode \texttt{PREVRANDAO}, \texttt{Geth/evm} triggers a runtime error, indicating an invalid memory address or nil pointer dereference. This issue is caused by an incorrect instruction set configuration, preventing the opcodes introduced in the post-Paris forks from executing correctly.
With our detailed reports, the developers confirmed and fixed this bug. As shown in Fig.~\ref{fig:case:geth}, developers fixed this bug by adding correct configurations for the supported instruction set from Line 8 to 10.

\begin{figure}
    \centering
\begin{lstlisting}[
        language=go,
        numbers=left,    
        numbersep=4pt,                   
        xleftmargin=1em,
        framexleftmargin=1em,
        morekeywords={NewInt,shanghaiTime,cancunTime}]
// ...
LondonBlock:                   new(big.Int),
ArrowGlacierBlock:             nil,
GrayGlacierBlock:              nil,
TerminalTotalDifficulty:       big.NewInt(0),
TerminalTotalDifficultyPassed: true,
// Bug fix: add the following lines to support newly introduced opcodes
MergeNetsplitBlock:            nil,
ShanghaiTime:                  &shanghaiTime,
CancunTime:                    &cancunTime
// ...
\end{lstlisting}
        \vspace{-0.15in}
    \caption{Case study \#2: the buggy implementation of \texttt{PREVRANDAO} in Geth, as well as the patch.}
    \vspace{-0.2in}
    \label{fig:case:geth}
    \Description{Case study \#2: the buggy implementation of \texttt{PREVRANDAO} in Geth, as well as the patch.}
\end{figure}

\answerbox{\textbf{\noindent Answer for RQ3: } On the one hand, the opcode-related bugs we found are widely present in the real world and can be triggered after setting appropriate parameters for EVMs. On the other hand, we found that these bugs can be triggered by a single opcode, highlighting the necessity of opcode-level testing.} 

\subsection{RQ4: Ablation Studies}
To assess the contribution of each component of our differential testing framework {\framework}, we performed ablation studies focusing on three key components: the seed generator (Seed), the control flow mutator (CFMut), and the argument mutator (ArgMut).
In a series of three independent experiments, we sequentially removed one component at a time and evaluated the impact on code coverage for both Geth and SealEVM.
Specifically, in the first experiment, we removed the seed from the seed generator along with its mnemonics (as shown in Fig.~\ref{fig:prompt:cfg}) and generated test inputs without them. 
In the second experiment, we removed the ICFG (as shown in Fig.~\ref{fig:prompt:cfg}) and disabled the control-flow-oriented mutation.
While in the third experiment, we disabled the argument mutator, using the raw output of LLMs as test inputs without mutation.
The impact of each component was assessed based on the observed code coverage.
As shown in Table~\ref{table:ablation}, the results demonstrated that each component positively contributed to the code coverage of the target EVMs. Removing any component led to a reduction in coverage.
Next, we will analyze the effect of each component. 
We first examined the impact of the seed generator. 
As shown in Table~\ref{table:ablation}, 
removing the seed generator resulted in the largest drop in code coverage (average code coverage loss of -14.96\%), followed by CFMut (-1.54\%) and ArgMut (-1.05\%).
We argue that the seeds and its mnemonics from seed generator provide a structured and well-defined example, enabling the LLMs to reason and generate meaningful test cases.
Without the ICFG, the LLMs failed to generate test cases that explore additional execution paths in the opcode definition, leading to a 1.54\% reduction in coverage.
Finally, we investigated the effect of the argument mutator. 
Since the argument mutator modifies arguments in the stack to generate random operands, disabling it restricted the exploration of corner cases in the specification, reducing the diversity of test inputs.

\begin{table}[t]
\caption{Code coverage results of ablation Study. "w/o" denotes "without," while "CFMut" and "ArgMut" refer to the control flow mutator and argument mutator, respectively.}
\label{table:coverage}
\resizebox{0.95\textwidth}{!}{%
\begin{tabular}{c|cc|cc|c}
\toprule
\textbf{Tool} & \textbf{Geth (All)} & \textbf{Geth (Opcode Funcs)} & \textbf{SealEVM (All)} & \textbf{SealEVM (Opcode Funcs)} & \textbf{Average Coverage Reduction}\\
\midrule
w/o~Seed & 48.30\% & 66.02\%  & 71.20\%  & 80.83\% & -14.96\% \\
w/o~CFMut   & 52.70\% & 89.50\%  & 88.60\%  & 89.27\% & -1.54\%\\
w/o~ArgMut   & 53.20\%  &  90.36\%  & 89.20\%  & 89.28\% & -1.05\% \\
\rowcolor{codegray} 
{\framework}  & 54.40\% & 91.85\%  & 90.10\% & 89.87\% & N/A \\ 
\bottomrule
\end{tabular}%
}
\vspace{-0.2in}
\end{table}

\answerbox{\textbf{\noindent Answer for RQ4: } Our ablation studies demonstrate that each component of {\framework} is essential for enhancing code coverage on target EVM implementations, underlining the necessity of integrating all three modules within {\framework}.}

\section{Discussion}
\label{sec:discussion}

\noindent \textbf{Ethical Consideration.}
Our differential testing has detected 26 distinct bugs from various EVMs, of which 22 have been confirmed or fixed by developers. To meet the ethical standards, we provide detailed bug reports and reproducible test inputs to developers once after the analysis is done. Moreover, all identified vulnerabilities are reported to both developers and relevant security databases, including the CNVD and NVD, for further disclosure and remediation.

\noindent \textbf{Limitation.} 
1) \textit{Test input generation.}
As we utilize LLM to generate test inputs, the main limitation lies on the LLM's restricted reasoning capabilities due to their intrinsic limitations. To solve this, we can introduce symbolic execution for a more accurate and comprehensive input generation, which is reserved for future work. To mitigate the potential unreliability of LLM's outputs caused by hallucinations, the self-refinement~\cite{madaan2024self} mechanism can be applied to iteratively enhance the consistency and correctness of the output.
2) \textit{Root cause localization.} Once we identified the responsible opcode, determining the underlying root cause for non-execution-stage bugs remains challenging, such as the bugs introduced at the EVM initialization stage. 
However, the localized responsible opcode can still aid the process by excluding irrelevant opcodes for the following manual inspection.
3) \textit{Opcode specification compatibility.}
Ethereum is evolving rapidly, which means that a new set of opcode specifications may be introduced in the future, like EOF \cite{EIP-3540}.
However, we underline that {\framework} is scalable.
By extracting the semantics of the corresponding specifications, we can equip LLMs with essential domain knowledge, enabling them to autonomously reason and generate effective test inputs.
This method ensures that {\framework} remains adaptable to the evolving landscape of Ethereum's development.

\section{Related Work}
\noindent \textbf{Blockchain Security.} Much of the existing research on blockchain security has focused on smart contract, with notable contributions from various tools and frameworks designed to identify weaknesses in smart contracts \cite{jiang2018contractfuzzer, shou2023ityfuzz, torres2021confuzzius, tsankov2018securify, GPTScan, EOSAFE, addressVer}. While substantial advances have been made in analyzing and detecting vulnerabilities in contract code, there is comparatively less focus on the security aspects of the foundational EVM. EVMFuzzer \cite{fu2019evmfuzzer} is the first differential testing tool for EVM. NeoDiff \cite{maier2021NeoDiff} utilizes bytecode-level generation test input generation method to fuzz the EVM. FuzzyVM \cite{fuzzyvm} and go evmlab \cite{goevmlab} are official differential testing tools maintained by Go-Ethereum Team. However, as introduced in section \S\ref{sec:challenge}, existing tools are limited in generating semantic-valid, diverse test inputs and lack an effective root cause analysis for identifying potential bugs. Besides, several studies have explored methods for testing blockchain consensus protocols ~\cite{yang2021fluffy,ma2023loki,chen2023tyr,bfttest} and RPC services ~\cite{kim2023etherdiffer, DBLP:conf/ndss/LiCLT0L21} in Ethereum execution clients. 

\noindent \textbf{Differential Testing.} Differential testing has been widely applied in both academia and industry, such as detecting issues in SSL/TLS implementations \cite{SSL2014,SSL2023,petsios2017nezha}, compilers \cite{guLLMBasedCodeGeneration2023} , runtime \cite{ cao2024wasmaker, zhou2023wadiff, li2023pyrtfuzz,bernhard2022jit, chen2016coverage, chen2019deep}. Specifically,  
PyRTFuzz designs a framework for Python runtime testing, which can generate applications covering runtime APIs for fuzzing. 
WASMaker \cite{cao2024wasmaker} uses real-world binaries to generate semantic-rich test inputs. 
WADIFF \cite{zhou2023wadiff} integrates symbolic execution techniques for the generation of test inputs. 
{\framework} is fundamentally inspired by existing differential testing methodologies and employs targeted testing approaches specifically designed for the unique context of the Ethereum Virtual Machine.

\section{Conclusion}
In this paper, we present {\framework}, a differential testing framework for Ethereum Virtual Machine (EVM) based on opcode-level test input generation.
By leveraging LLM's reasoning and comprehension capabilities and static analysis methods, {\framework} can generate semantically valid and diverse test inputs according to the specifications.
To assist following debugging processes, we also propose an automated bug identification and root cause localization algorithm.
Compared with state-of-the-art baselines, our evaluation illustrates that {\framework} can achieve up to 410.8\% code coverage improvement. 
Among nine EVM implementations spanning multiple scenarios, we have uncovered 26 previously unknown real-world bugs, 22 of which have been confirmed or fixed and three of them have been assigned CNVD IDs.
Additionally, we investigated that 7.21\% real-world smart contracts can trigger these bugs under certain environmental settings.

\section*{Data Availability}
The dataset, artifact and the main experimental results of {\framework} are available at \cite{code}.

\bibliographystyle{ACM-Reference-Format}
\bibliography{references}

\end{document}